\documentclass[twocolumn]{aastex702}
\usepackage{hyperref}
\usepackage{color}
\usepackage{amssymb}
\usepackage{amsmath}
\usepackage[ruled,linesnumbered]{algorithm2e}
\usepackage{footmisc}
\usepackage{ulem}

\shorttitle{Ca Evolution in High-z SNe Ia}
\shortauthors{Chen et al.}
\begin{document}

\newcommand{\AIAItwo}{\citep[hereafter AIAI2]{Chen2024AIAI2}}
\defcitealias{Chen2024AIAI2}{AIAI2}

\title{High-Redshift Type Ia Supernovae Exhibit Enhanced Calcium Abundances}
\correspondingauthor{Ulisses Braga-Neto}
\email{ulisses@tamu.edu}

\author[0000-0003-3021-4897]{Xingzhuo Chen}
\affiliation{Texas A\&M Institute of Data Science,
Texas A\&M University,
College Station, TX, USA}
\email{chenxingzhuo@tamu.edu}

\author[0000-0002-1210-2173]{Ulisses Braga-Neto}
\affiliation{Department of Electrical and Computer Engineering, Texas A\&M University, College Station, TX, USA}
\email{ulisses@tamu.edu}

\author{Lifan Wang}
\affiliation{George P. and Cynthia Woods Mitchell Institute for Fundamental Physics \& Astronomy, \\
Texas A. \& M. University, Department of Physics and Astronomy, 4242 TAMU, College Station, TX 77843, USA}
\email{lifan@tamu.edu}

\begin{abstract}

Type Ia supernovae (SNe Ia) are major contributors to cosmic chemical enrichment, and their elemental abundances provide a probe of progenitor properties and explosion physics across cosmic time. 
We employ an artificial intelligence-assisted inversion technique to analyze spectra of high-redshift SNe Ia from the Supernova Legacy Survey and of gravitationally lensed SNe Ia observed by the James Webb Space Telescope, extending the sample to redshift 2.05. 
We find a positive correlation between SN Ia calcium abundance and redshift. 
The redshift-dependent variation in calcium abundance exceeds that predicted by SN Ia nucleosynthesis simulations with varying progenitor metallicities, suggesting that high-redshift SNe Ia may undergo different explosion mechanisms from nearby SNe Ia. 


\end{abstract}

\keywords{Type Ia supernovae, Radiative transfer simulations, Convolutional neural networks, Microlensing, Chemical evolution}

\section{Introduction}\label{sec:intro}

Type Ia supernovae (SNe Ia) have long served as standardizable candles in cosmology, playing a pivotal role in the discovery of the accelerating expansion of the universe and the precise measurement of dark energy \citep{Phillips1993Relation, Abbott2019Cosmology, Brout2022Pantheon}. 
Beyond their utility in measuring cosmic distances, SNe Ia are also major contributors to the chemical enrichment of the universe, particularly synthesizing a significant fraction of iron-peak and intermediate-mass elements. 
As the end-points of carbon-oxygen white dwarfs in binary systems, the nucleosynthesis yields of SNe Ia are highly sensitive to the progenitor metallicity and the details of the explosion mechanism (e.g., deflagration versus delayed-detonation) \citep{Gamezo2004DDT, Seitenzahl2013N100, Shen2018NewModels}. 
Theoretical nucleosynthesis simulations have provided crucial insights into how different progenitor metallicities influence the explosion dynamics and elemental yields in SNe Ia. 
Detailed three-dimensional delayed-detonation models \citep{Seitenzahl2013N100, Travaglio2004Nucleo} demonstrate that lower metallicity progenitors tend to produce more complete silicon burning, resulting in higher calcium yields in the outer ejecta. 
This effect is primarily attributed to reduced opacity in lower metallicity environments, which allows for more efficient energy transfer and a more deflagration-to-detonation transition \citep{Reinecke2002LEAFS}. 
Additionally, sub-Chandrasekhar-mass white dwarf models \citep{Shen2018NewModels,Keegans2023Nucleosynthesis} suggest that lower metallicity companion stars may lead to different ignition conditions and potentially enhanced calcium production. 
These findings are consistent with the observed trend of higher calcium abundances in high-redshift SNe Ia, which may reflect the systematically lower progenitor metallicities in the early universe \citep{Kromer2017Hesma}. 
Consequently, the elemental abundances derived from SN Ia spectra offer a unique window into both stellar evolution across cosmic time and the robustness of SNe Ia as cosmological probes. 

Observationally, calcium spectral lines (e.g., the Ca~II near-infrared triplet and the Ca~II H\&K lines) often exhibit complex kinematic and geometrical properties; for instance, spectropolarimetry has revealed high-velocity, clumpy calcium structures distributed asymmetrically in the outer ejecta, suggesting a deep connection between calcium synthesis and the explosion geometry \citep{Wang_2003, Yang_2022, Wang2008SpecPol, Hoeflich2023SN2019np}. 
In the local universe, the observed calcium abundances in SN Ia spectra exhibit a complex interplay with light-curve shape and velocity, yet the environmental dependence—particularly on global metallicity—remains highly debated. 
To disentangle the progenitor metallicity effects from intrinsic explosion physics, one must trace the evolution of SN Ia elemental abundances over a wide redshift baseline. 
However, such endeavors have been historically hampered by the severe signal-to-noise ratio (SNR) degradation of optical spectra at $z > 1$, rendering traditional abundance analysis techniques unfeasible for distant objects. 

The Supernova Legacy Survey (SNLS) \citep{Astier2006SNLS} has provided an extensive and homogeneous sample of SNe Ia up to $z \approx 1$, offering a rich archival dataset of spectral observations. 
More recently, the advent of the James Webb Space Telescope (JWST) has revolutionized the study of the high-redshift transient universe. 
In particular, the two high-redshift SNe Ia SN~2025ogs \citep{Siebert2026SN2025ogs} and SN~H0pe \citep{Chen2024SNH0pe}, observed by JWST, have enabled spectroscopic observations of SNe Ia at $z > 2$ for the first time. 
The synergy between the statistical power of the SNLS sample and the deep, high-redshift reach of JWST lensed SNe Ia presents an unprecedented opportunity to map the chemical evolution of SNe Ia from the local universe to the epoch of $z > 2$.

Despite the emergence of new observational data, extracting precise elemental abundances from a large, heterogeneous, and often low-SNR spectroscopic dataset remains a formidable computational challenge. 
Traditional spectral fitting methods, which typically rely on radiative transfer codes (e.g., \citealt{Kasen2006Sedona, Kerzendorf2014Tardis, Kromer2009ARTIS, Hillier2012CMFGEN}), are computationally expensive in Bayesian analyses and susceptible to degeneracies, making them impractical for large-scale high-redshift surveys. 
To overcome this bottleneck, the Artificial Intelligence-Assisted Inversion (AIAI) framework was established \citep{Chen2020AIAI}. 
By training convolutional neural networks (CNNs) on large-scale synthetic spectra generated from radiative transfer simulations, the AIAI method demonstrated that deep learning can efficiently capture the complex, non-linear mappings between spectral features and physical parameters, enabling rapid and robust abundance inference. 
This framework was subsequently advanced in \AIAItwo, which specifically targeted the $^{56}$Ni properties in SNe Ia in a longer observation time horizon. 
By quantifying the spectral features associated with $^{56}$Ni, the updated AIAI model proved capable of disentangling intricate spectral degeneracies and extracting reliable physical parameters even from noisy observations, establishing a powerful and mature pipeline for quantitative SN Ia spectroscopy. 

In this paper, we leverage deep learning algorithms to analyze the spectral observations of SNe Ia from the SNLS dataset alongside recent JWST observations of gravitationally lensed, high-redshift SNe Ia. 
By establishing a unified framework for abundance inference across a wide redshift range, we systematically investigate the cosmic evolution of SN Ia nucleosynthesis. 
Strikingly, our analysis reveals a robust positive correlation between the calcium abundance in SNe Ia and redshift. 
This finding suggests significant evolution in the progenitor properties or explosion conditions of SNe Ia over cosmic time, potentially reflecting the decreasing average host-galaxy metallicity toward earlier epochs, which favors more complete silicon burning and consequently higher calcium yields in the outer ejecta.

The structure of this paper is as follows. 
Section~\ref{sec:data} describes the SNLS and JWST spectroscopic datasets used in this study. 
Section~\ref{sec:method} details the architecture and training of our deep learning abundance inference model. 
Section~\ref{sec:results} presents the derived calcium abundances and their correlation with redshift. 
Section~\ref{sec:discussion} discusses the physical implications of our findings in the context of SN Ia progenitor models and cosmic chemical evolution. 

\section{Data}\label{sec:data}

In \citetalias{Chen2024AIAI2}, the AIAI method was applied to 124 SNe Ia (including Ia-91T and Ia-91bg subtypes) to estimate ejecta elemental abundances. 
These SNe Ia were reported by several supernova survey programs (e.g., Kaepora \citealt{Siebert2019Kaepora}), and the highest-redshift object is at $z=0.0835$. 
To extend this low-redshift sample, we include SNe Ia from the completed SNLS program and from ongoing JWST programs that survey lensed SNe Ia. 

\subsection{SNLS}

The Supernova Legacy Survey (SNLS) is a major astronomical survey project designed to detect and study Type Ia supernovae (SNe Ia) over a wide range of redshifts \citep{Astier2006SNLS}. 
The survey was conducted using the Canada-France-Hawaii Telescope (CFHT) and its wide-field camera, MegaCam, covered four fields of 1 square degree each \citep{Sullivan2006SNLSselect}. 
The survey operated from 2003 to 2008, collecting photometric data for $\sim$472 SNe Ia up to redshift $z\sim 1.1$ in the $g'r'i'z'$ filters every 3--5 days to enable SN Ia candidate selection, light-curve fitting, and follow-up spectroscopic observations. 

The spectroscopic follow-up observation of SNLS is performed by several telescopes, and the reduced spectra from the Very Large Telescope (VLT); the Keck telescope; the Gemini North and South telescope are summarized in \cite{Sullivan2011SNLS3} and are available to download\footnote{\href{https://utoronto.scholaris.ca/items/d134a48b-82d0-46db-a805-5e9362d7b45e}{https://utoronto.scholaris.ca/items/d134a48b-82d0-46db-a805-5e9362d7b45e}}. 
Reported in \cite{Balland2009SNLSVLT}, 139 spectra from 124 SNe Ia are observed with FORS1 or FORS2 facilities at the VLT, the reduced spectra with removed host galaxy contamination covering the observer-frame wavelength between 4000 $\rm\AA$ and 9000 $\rm\AA$ are publicly available through the SNLS3 archive \citep{Sullivan2011SNLS3}. 
Moreover, 36 SNe Ia are observed by the LRIS facility on the Keck I telescope, the reduced spectra with removed host galaxy contamination covering the observer-frame wavelength between 3400 $\rm\AA$ and 8300 $\rm\AA$ are reported in \cite{Ellis2008SNLSKeck}. 

The SN Ia candidates observed with the Gemini North and South telescopes were observed using the GMOS facilities, which cover observer-frame wavelengths from 5000 to 9500 $\rm\AA$. 
The spectra are reported in multiple papers with different light-curve fitting methods and host galaxy contamination removal techniques. 
For example, \cite{Howell2005Gemini} reported 34 SNe Ia spectra observed by the Gemini North or Gemini South telescope, and the corresponding rest-frame phase of SNe Ia spectra determined from light-curve and spectral fitting methods, while \cite{Bronder2008Gemini} reported slightly different values of the rest-frame phase of the same group of SNe Ia. 
Moreover, both \citet{Howell2005Gemini} and \citet{Bronder2008Gemini} used a $\chi^2$ fitting method to determine the host-galaxy contamination, whereas \citet{Walker2011Gemini} adopted an approach similar to that of \citet{Ellis2008SNLSKeck} to remove it. 

In this research, we adopt the following data: 

\begin{itemize}
    \item 139 spectra of 124 SNe Ia are observed by the VLT. The spectra after removing the host galaxy contaminations and extinction effects are reported in \cite{Balland2009SNLSVLT}. 
    \item 36 spectra of 36 SNe Ia are observed by the Keck telescope. The spectra after removing host galaxy contaminations and host galaxy extinction are reported in \cite{Ellis2008SNLSKeck}. 
    \item 34 spectra of 34 SNe Ia are observed by the Gemini North or Gemini South telescope, reported in Table 2 of \cite{Howell2005Gemini}. We adopt the rest-frame phase of SNe Ia spectra determined from the spectral fitting method. 
    \item 53 spectra of 53 SNe Ia are observed by the Gemini North or Gemini South telescope, reported in Table A.3 of \cite{Bronder2008Gemini}. Note that these SNe Ia are subject to less than 65\% contaminations from the host galaxy spectra. 
\end{itemize}

Twenty-five spectra are included in both Gemini data sets; the only difference is that their spectral phases were derived using different methods. 
A detailed list of the SN Ia spectra used in this work is provided in Appendix \ref{sec:sntable}. 

\subsection{JWST Supernovae}

SN~2025ogs is a spectroscopically normal SN Ia at redshift $z=2.05\pm 0.01$ \citep{Siebert2026SN2025ogs}. 
SN~2025ogs was observed spectroscopically with the JWST NIRSpec facility on May 6, 2025. 
The spectrum was compared with spectral templates in the Kaepora spectroscopic database \citep{Siebert2019Kaepora}, and the light curves observed with JWST NIRCam were fitted using BayeSN \citep{Mandel2022Bayesn}. \citet{Siebert2026SN2025ogs} reported a spectroscopic phase of $\sim$+1 day and negligible host-galaxy dust extinction. 

SN~H0pe is a spectroscopically normal SN Ia at redshift $z=1.78\pm 0.01$ \citep{Pascale2025SNH0pe}. 
On April 22, 2023, SN~H0pe was observed with the JWST NIRSpec facility using multi-object spectroscopy, yielding three gravitationally lensed SN spectra at different phases simultaneously \citep{Chen2024SNH0pe}. 
Using SALT3-NIR \citep{Pierel2022SALT3NIR}, the spectra were assigned phases of $\sim$+4.2, +26.0, and +45.1 days, respectively. 
Because \citetalias{Chen2024AIAI2} covers phases between $-10$ and +20 days, we adopt the SN~H0pe spectrum at +4.2 days in this work. 
Using the SN Ia template of \citet{Hsiao2007Template} for spectral fitting, we find a host-galaxy extinction of $E(B-V)=0.27\pm 0.02$ for SN~H0pe. 


\section{Methods}\label{sec:method}

The artificial intelligence-assisted inversion (AIAI) method introduced by \citet{Chen2020AIAI} and \citetalias{Chen2024AIAI2} uses multi-residual convolutional neural networks to predict SN Ia ejecta structures from optical spectra. 

First, we use the one-dimensional, time-independent Monte Carlo radiative-transfer program TARDIS \citep{Kerzendorf2014Tardis} to simulate optical spectra for a series of SN Ia ejecta structures drawn from a fiducial distribution. 
TARDIS assumes a one-dimensional, spherically symmetric ejecta structure and blackbody radiation at the inner boundary as the optical energy source. It solves the plasma level populations using the local thermodynamic equilibrium (LTE) approximation with a dilution factor \citep{Mazzali1993DiluteLte}. 
We divide the one-dimensional ejecta structure radially into six zones to distinguish the spatial distributions of different elements. 
From the innermost to the outermost zone, the velocity ranges are 3000--7000, 7000--10,000, 10,000--13,000, 13,000--17,000, 17,000--24,000, and 24,000--35,000 km s$^{-1}$ for Zones 0--5, respectively. 

Subsequently, we train a suite of multi-residual convolutional neural networks \citep{Abdi2016MRNN} to predict SNe Ia ejecta structures from optical spectrum inputs using TARDIS simulation data. 
We adopt deep ensemble neural networks \citep{Laksh2016DeepEnsemble} to predict both the mean values and the $1\sigma$ model uncertainties of the SN Ia elemental mass fractions, assuming a Gaussian error distribution. 

Third, the AIAI method is applied to the observed spectra of SNe Ia to infer the SNe Ia ejecta structure. 
The TARDIS simulations are performed on the grid of predicted ejecta structures, in order to select the optimal ejecta models that fit the observed spectral time sequence of each SNe Ia. 
This spectral-fitting method constrains spectra of the same SN Ia at different phases to a common density profile, assuming that the SN Ia ejecta undergo homologous expansion after the explosion. 
A detailed introduction of the spectral fitting method is presented in Section 2.4 of \citetalias{Chen2024AIAI2}. 

As a validation criterion for the spectral-fitting results and elemental-abundance estimates, we calculate the spectral mean squared error (MSE) for each SN Ia: 

\begin{equation}\label{eq:mse}
    \rm{MSE}=\sum_\lambda \left( F_{\rm obs}(\lambda)-F_{\rm sim}(\lambda) \right)^2/n \ ,
\end{equation}
where $F_{\rm obs}$ and $F_{\rm sim}$ are the observed and simulated spectra, respectively, normalized by their corresponding means, and $n$ is the number of wavelength bins. 
The wavelength grid is resampled according to the simulated spectrum, and the regions with missing observations are set to have zero weight. 

\begin{figure*}
    \centering
    \includegraphics[width=0.49\textwidth]{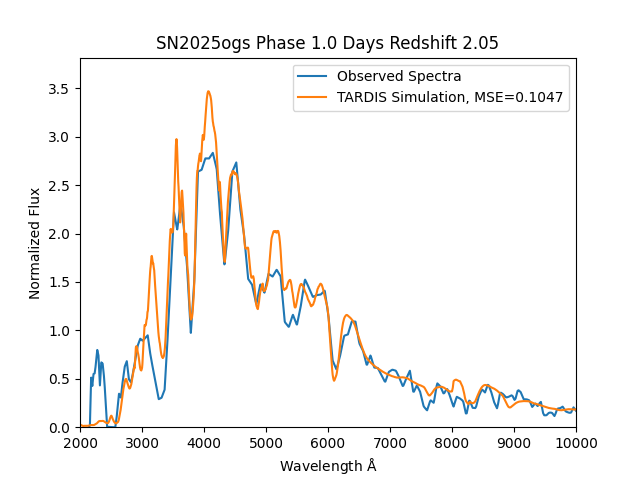}
    \includegraphics[width=0.49\textwidth]{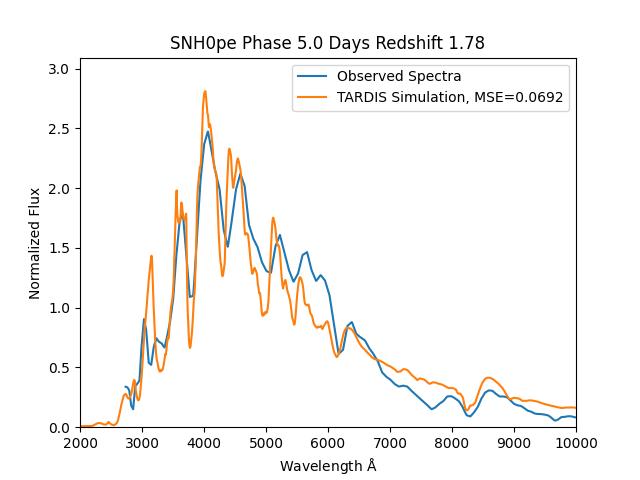}
    \caption{The JWST observed spectra and the AIAI fitted spectra of SN 2025ogs (left) and SN H0pe (right). }
    \label{fig:SEDONA_2025ogs}
\end{figure*}

Figure \ref{fig:SEDONA_2025ogs} shows the JWST observed spectra and the AIAI fitted spectra of SN~2025ogs (left) and SN~H0pe (right). 
A sample of the SNLS spectra, which has the smallest, median, and largest MSE values, are shown in Appendix \ref{sec:snspectra}, in Figure \ref{fig:SNLSKeck}, Figure \ref{fig:SNLSVLT}, Figure \ref{fig:SNLSGemini}, Figure \ref{fig:SNLSGemini2}. 

\section{Results}\label{sec:results}

\subsection{Elemental Abundances}

\begin{figure*}
    \includegraphics[width=0.33\textwidth]{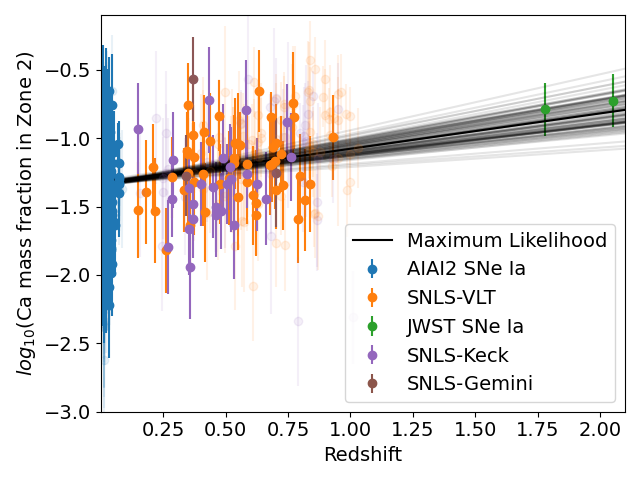}
    \includegraphics[width=0.33\textwidth]{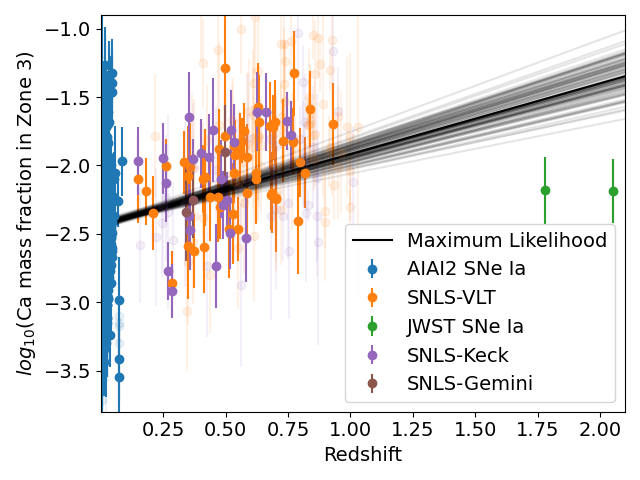}
    \includegraphics[width=0.33\textwidth]{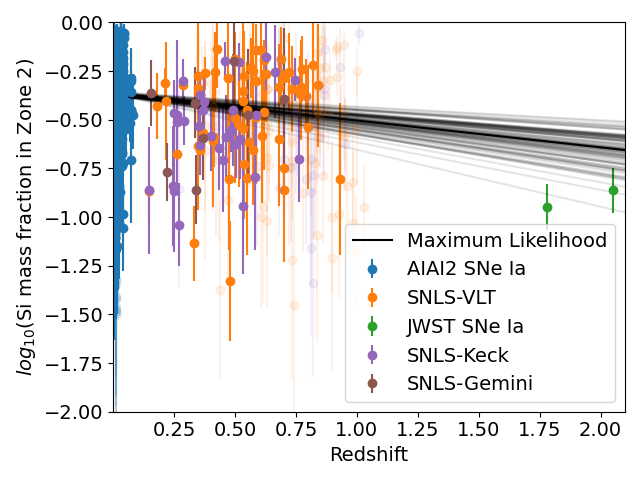}
    \includegraphics[width=0.33\textwidth]{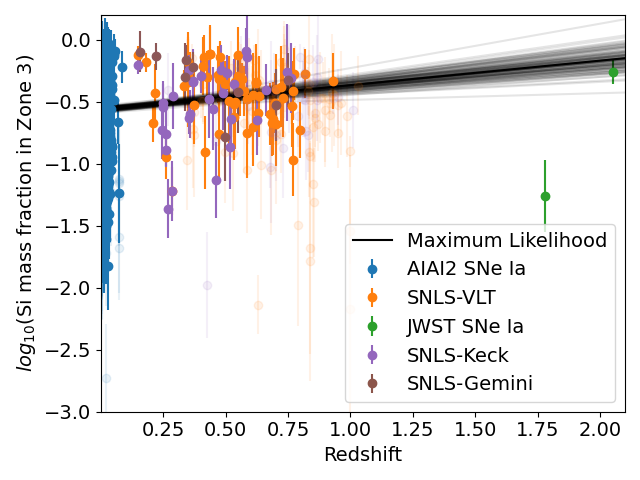}
    \includegraphics[width=0.33\textwidth]{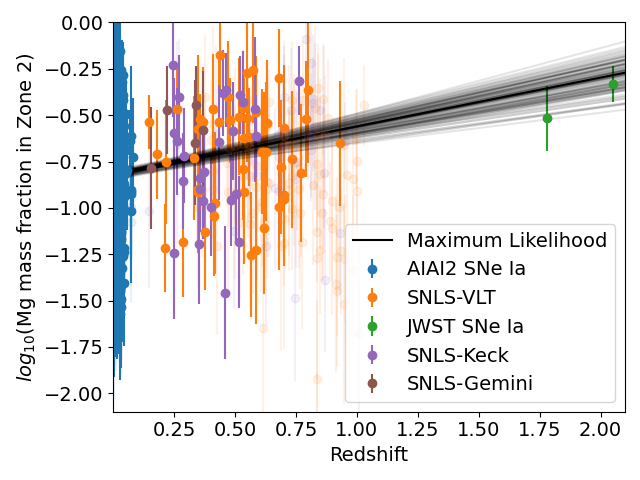}
    \includegraphics[width=0.33\textwidth]{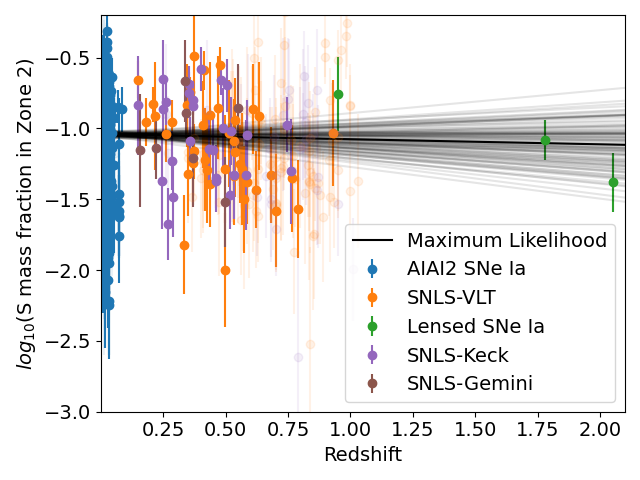}
    \caption{The correlation between redshift and elemental mass fractions of (upper left) Ca in Zone 2, (upper middle) Ca in Zone 3, (upper right) Si in Zone 2, (lower left) Si in Zone 3, (lower middle) Mg in Zone 2, (lower right) S in Zone 2. The maximum likelihood estimation results are shown in black lines with 100 samples from the posterior distribution. The SNe Ia spectra not selected for maximum likelihood estimate (due to model uncertainty larger than 0.4 or spectral fitting MSE larger than 0.2) are transparent. }\label{fig:ElemZ}
\end{figure*}

We assume the elemental abundances and redshifts follow a linear correlation of the form:

\begin{equation}\label{eq:linear}
    log_{10}(E)=m z + b \ , 
\end{equation}
where $E$ is the selected elemental mass fraction in a zone, $z$ is redshift, and $m$ and $b$ are the model parameters estimated by maximum likelihood. 
The logarithmic maximum likelihood function is: 

\begin{equation}\label{eq:MLE}
    ln(L)=-\frac{1}{2} \sum_n \left( \frac{(log_{10}(E_n)-m z_n-b)^2}{s_n^2}+ln(2\pi s_n^2)\right) \ , 
\end{equation}
where $s_n^2=\sigma_n^2+f^2(m z_n+b)^2$ is the total uncertainty, accounting for both the elemental-abundance model uncertainty ($\sigma_n$) provided by the deep ensemble network and the uncertainty-underestimation factor $f$; $log_{10}(E_n)$ is the elemental abundance of the $n$th SN Ia; and $z_n$ is the redshift of the $n$th SN Ia. 
We use the \texttt{emcee} \citep{Foreman2013Emcee} Markov chain Monte Carlo program to obtain maximum-likelihood estimates of the parameters $m$, $b$, and $f$. 
For the maximum likelihood estimation, we select only those SNe Ia with spectral fitting MSE values (Equation \ref{eq:mse}) below 0.2 and deep-learning predictive uncertainties below 0.4 dex. 
In Figure \ref{fig:ElemZ}, we show the elemental abundances of Ca in Zone 2, Ca in Zone 3, Si in Zone 2, Si in Zone 3, Mg in Zone 2, and S in Zone 2. 
The maximum-likelihood estimates for all elements and zones are available in the Zenodo online material \footnote{\href{https://zenodo.org/uploads/21877558}{https://zenodo.org/uploads/21877558}}. 

\subsection{Comparison to Nucleosynthesis Results}

Several nucleosynthesis simulations of different SN Ia explosion scenarios \citep{Seitenzahl2013N100,Keegans2023Nucleosynthesis} assume that progenitor metallicity does not affect the explosion hydrodynamics and use $^{22}$Ne as a proxy for progenitor metallicity. 
Across the different explosion mechanisms, the nucleosynthesis results suggest that increasing progenitor metallicity increases the yields of V and Mn and decreases those of Mg, S, Ar, Ca, and Ti. The Si and Cr yields do not change significantly, whereas iron-group elements show isotope-dependent trends. 

We select the Ca/Si ratios in Zone 2 (10,000--13,000 km s$^{-1}$) and Zone 3 (13,000--17,000 km s$^{-1}$), the S/Si ratio in Zone 2, and the Mg/Si ratio in Zone 2 to compare the predicted elemental abundances with the nucleosynthesis simulations for the following reasons: 

\begin{itemize}
    \item According to SNe Ia hydrodynamic simulations (e.g. \cite{Seitenzahl2013N100}) and observational results (e.g. \citetalias{Chen2024AIAI2}), the intermediate-mass elements Si, S, Ca are concentrated in the velocity range 10000 - 17000 km/s. 
    \item Silicon has the largest mass fraction in the 10,000--17,000 km s$^{-1}$ velocity range, and its abundance is estimated accurately by AIAI. It can therefore serve as a reference for comparing the mass fractions of Mg and Ca across different SN Ia progenitor masses and density profiles. 
    \item We exclude iron-group elements because a large fraction of these elements is synthesized below 10,000 km s$^{-1}$ and is not observable in most available high-redshift spectra near $B$-band maximum light. 
    \item More than 50\% of the SNLS spectra have predictive uncertainties below 0.4 dex for the selected elemental abundances and zones, whereas other elements and zones have large predictive uncertainties. 
\end{itemize}

\begin{figure*}
    \centering
    \plottwo{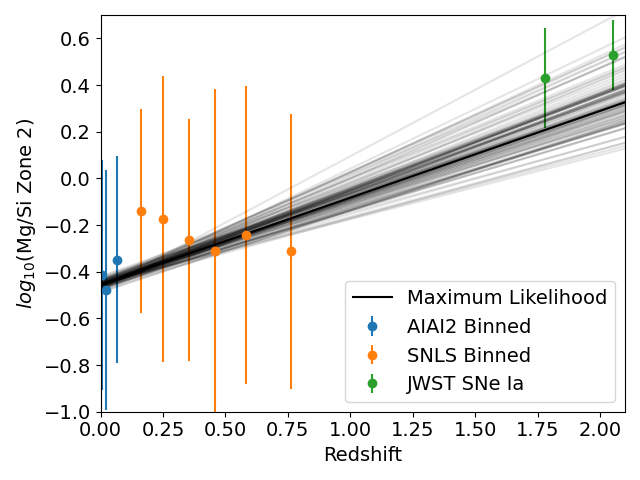}{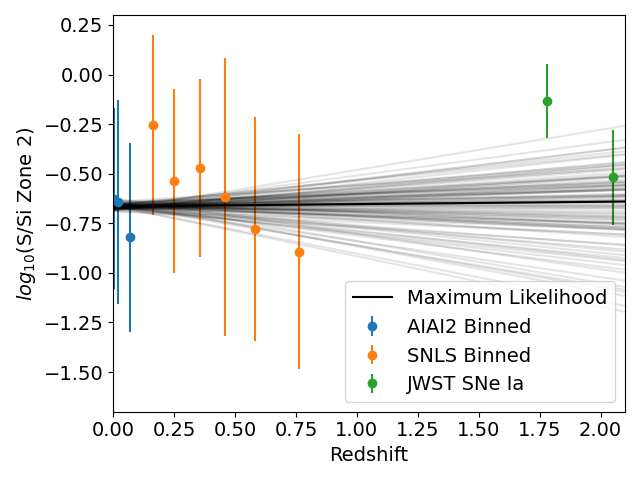}
    \plottwo{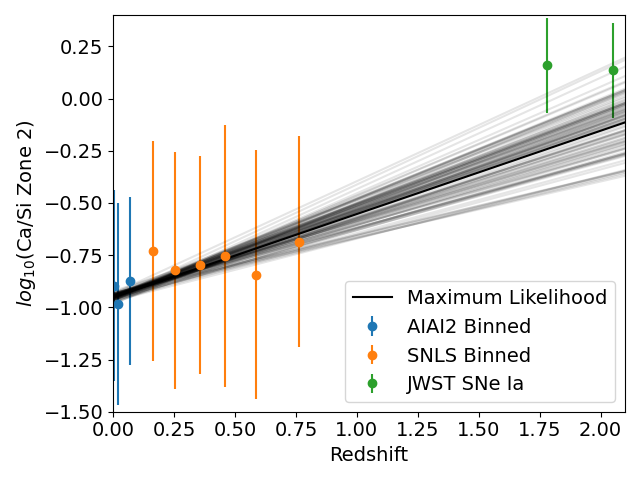}{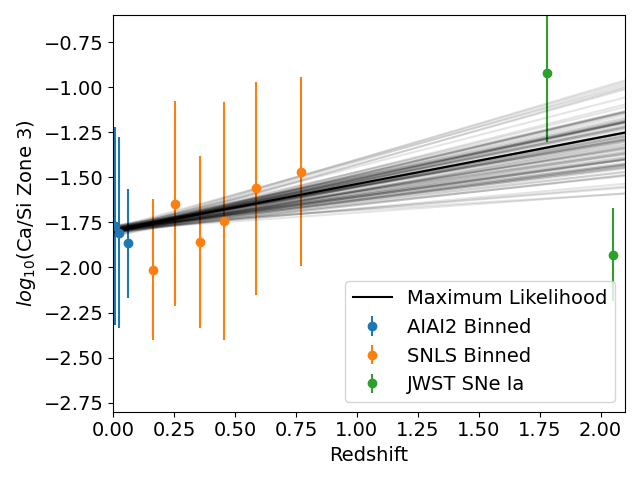}
    \caption{The correlation between the elemental abundance ratios and redshift. Upper Left: the Mg/Si elemental abundance ratio in Zone 2 (10000-13000 km/s). Upper Right: the S/Si elemental abundance ratio in Zone 2 (10000-13000 km/s). Lower Left: the Ca/Si elemental abundance ratio in Zone 2. Lower Right: the Ca/Si elemental abundance ratio in Zone 3. The low-redshift SNe Ia from \citetalias{Chen2024AIAI2} are averaged into 3 bins with respect to the redshift, marked in blue. The SNLS SNe Ia from VLT, Keck, and Gemini telescopes are averaged into 7 bins with respect to the redshift, marked in orange. The lensed SNe Ia are marked in green. }
    \label{fig:ElRatio}
\end{figure*}

Figure \ref{fig:ElRatio} shows the evolution of Mg/Si, S/Si, and Ca/Si element ratios in Zone 2 or Zone 3 with redshift. 
We find positive correlations between redshift and S/Si in Zone 2, Ca/Si in Zone 2, and Ca/Si in Zone 3, whereas no distinct correlation is observed for Mg/Si in Zone 2. 
We adopt a similar maximum likelihood estimation method as illustrated in Equation \ref{eq:linear} and Equation \ref{eq:MLE} to fit a linear correlation between the redshift and element ratios, then calculate the element ratio differences between redshift 0 and redshift 2. 
In Table \ref{tab:ELRatio}, we compare the observational element ratio differences with the results from nucleosynthesis simulations. 
For comparison, we adopt an $1.4M_{\odot}$ deflagration-to-detonation model (T1.4), an $1.0M_{\odot}$ double-detonation model (S1.0), and a $0.8M_{\odot}$ double-detonation model (M0.8) from \citet{Keegans2023Nucleosynthesis}, as well as a $1.4M_{\odot}$ delayed-detonation model (N100) from \citet{Seitenzahl2013N100}. 
Note that the progenitor metallicity of the nucleosynthesis simulations at redshift 2 is assumed to be $Z=0.01Z_{\odot}$, which is significantly lower than the values reported in galaxy spectroscopic survey programs (e.g. \cite{Halliday2008GMASS} report that the cosmic metallicity at redshift 2 should be $log_{10}(Z/Z_\odot)=-0.574\pm0.159$). 
We notice that the evolution of SNe Ia Mg/Si and Ca/Si element ratios observed with redshift is much more significant than the nucleosynthetic contributions from the progenitor metallicities, and the S/Si element ratio in both the observations and nucleosynthesis simulations are not changing significantly with redshift. 

Because the prominent Mg spectral line, Mg II $\lambda 4481$, in SNe Ia are blended with iron-group elements, and only Mg in Zone 2 are successfully measured among most of the SNLS SNe Ia, the Mg elemental abundances measured in this work may not be the optimal tracer of elemental abundance evolution with redshift. 
In comparison, the Ca spectral lines, including the Ca II H\&K lines and the Ca II infrared triplet lines are observable in most of the SNe Ia, and the Ca elemental abundances in both Zone 2 and Zone 3 are measured in most of the SNLS SNe Ia. 
Therefore, Ca is the most suitable tracer of elemental-abundance evolution with redshift in this analysis. 

\begin{table}
    \centering
    \begin{tabular}{c|ccccc}
        \hline
        $log_{10}\left(\frac{(\frac{E_1}{E_2})_{z=2}}{(\frac{E_1}{E_2})_{z=0}}\right)$ & Observed & T1.4 & S1.0 & M0.8 & N100 \\
        \hline
        \hline
        Mg/Si Zone 2 & 0.76$\pm$0.09 & 0.403 & 0.676 & 0.473 & 0.273 \\
        S/Si Zone 2  & 0.03$\pm$0.16 & 0.031 & 0.032 & 0.054 & 0.008 \\
        Ca/Si Zone 2 & 0.79$\pm$0.12 & 0.077 & 0.113 & 0.235 & 0.073 \\
        Ca/Si Zone 3 & 0.50$\pm$0.14 & 0.077 & 0.113 & 0.235 & 0.073 \\
        \hline
    \end{tabular}
    \caption{The logarithmic element ratio at redshift 2 divided by the element ratio at redshift 0. }
    \label{tab:ELRatio}
\end{table}

\subsection{Host Galaxy Properties}

In this section, we identify SN Ia host galaxies in the Dark Energy Spectroscopic Instrument (DESI) Data Release 1 value-added catalog \citep{Desi2026Desi} and the Sloan Digital Sky Survey (SDSS) Data Release 16 galaxy catalog \citep{Comparat2017SDSS} to investigate correlations between SN Ia elemental abundances and host-galaxy properties. 
We use the following criteria to identify the host galaxy: 

\begin{itemize}
    \item The angular separation between the SN Ia and the host galaxy should be smaller than 0.1 degree. 
    \item The redshift difference between the SN Ia and the host galaxy should be smaller than 0.002. 
    \item If multiple galaxies satisfy the first two criteria, then the galaxy with the smallest redshift difference is identified as the host galaxy. 
\end{itemize}

Among the 124 SNe Ia listed in \citetalias{Chen2024AIAI2}, 98 have an identified host galaxy in the DESI catalog and 109 have an identified host galaxy in the SDSS catalog. 
Among the 201 SNe Ia observed by SNLS and adopted in this work, 80 have an identified host galaxy in the DESI catalog and 41 have an identified host galaxy in the SDSS catalog. 
The host galaxy properties from the DESI catalog are estimated using the STARLIGHT galaxy spectral synthesis code \citep{Cid2011Starlight}. 
The host-galaxy properties in the SDSS catalog are estimated using the FIREFLY spectral-synthesis program \citep{Wilkinson2017Firefly}, which adopts the initial mass function of \citet{Chabrier2003IMF} and the MILES spectral library of \citet{Falcon2011Miles}. 
The host galaxies of the two JWST SNe Ia are not included in these catalogs. We adopt host-galaxy stellar masses of $log_{10}(M_*/M_\odot)=8.6\pm0.1$ for SN~2025ogs \citep{Siebert2026SN2025ogs} and $log_{10}(M_*/M_\odot)=11.69\pm0.01$ for SN~H0pe \citep{Frye2024SNH0pe}. 

\begin{figure*}
    \centering
    \includegraphics[width=0.32\textwidth]{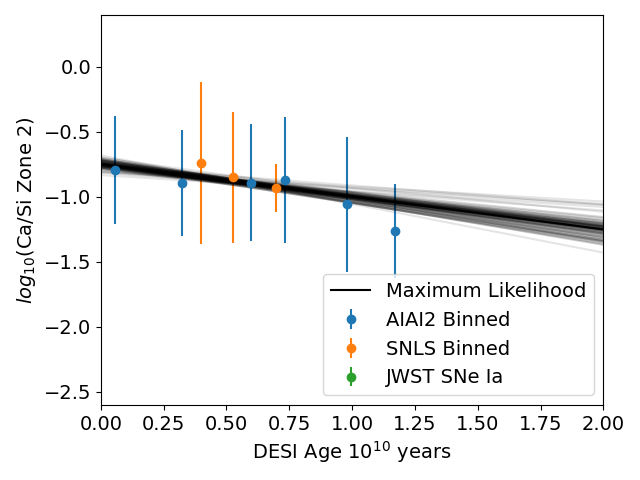}
    \includegraphics[width=0.32\textwidth]{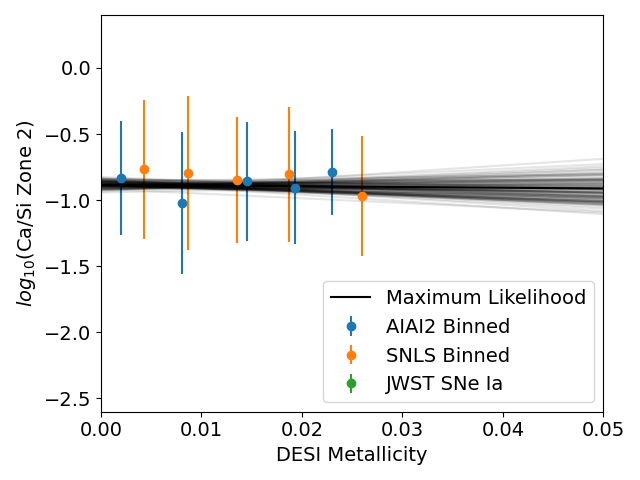}
    \includegraphics[width=0.32\textwidth]{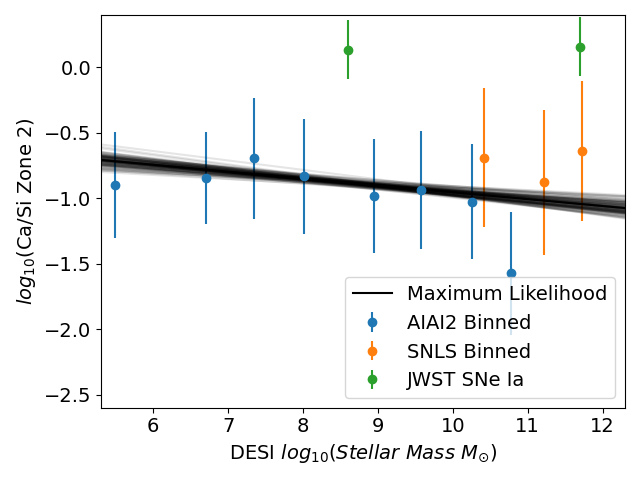}
    \includegraphics[width=0.32\textwidth]{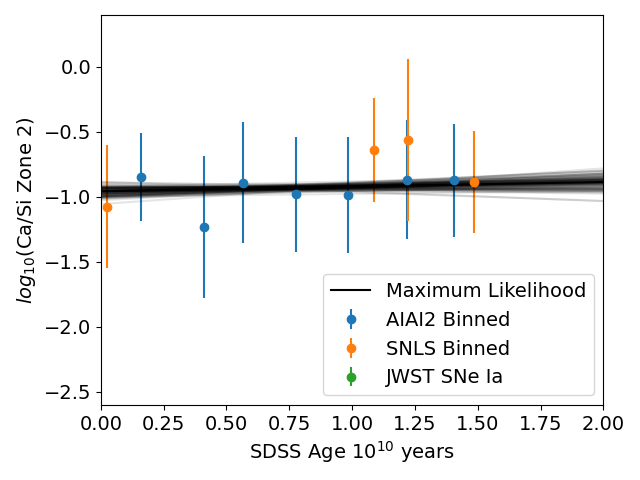}
    \includegraphics[width=0.32\textwidth]{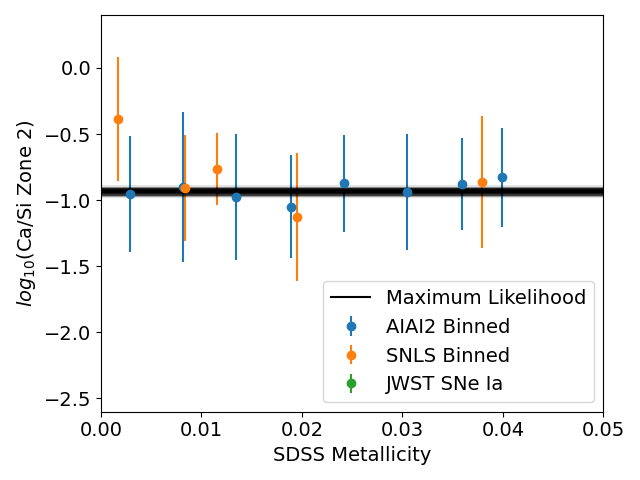}
    \includegraphics[width=0.32\textwidth]{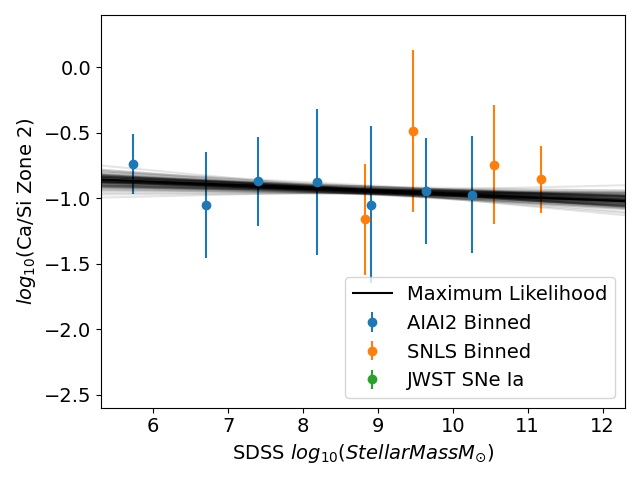}
    \includegraphics[width=0.32\textwidth]{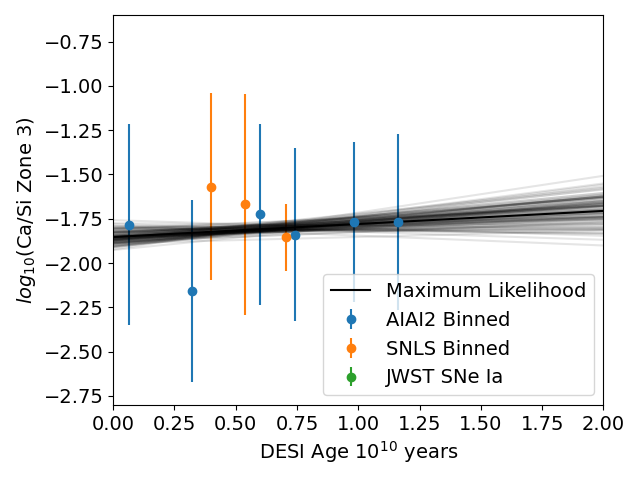}
    \includegraphics[width=0.32\textwidth]{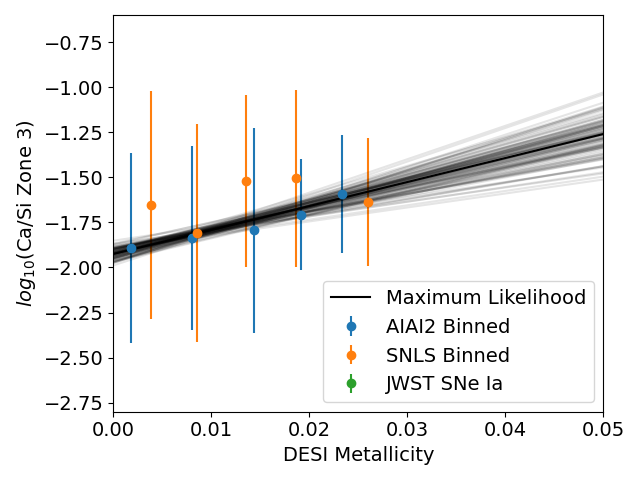}
    \includegraphics[width=0.32\textwidth]{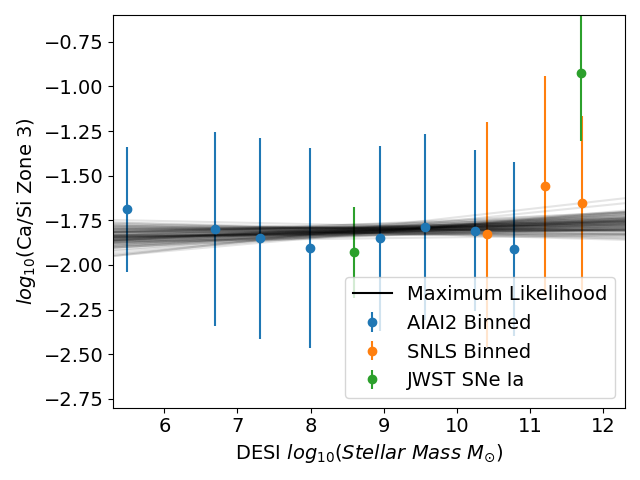}
    \includegraphics[width=0.32\textwidth]{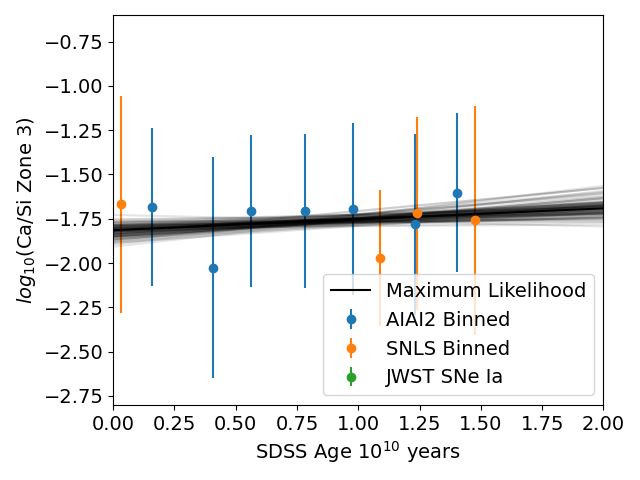}
    \includegraphics[width=0.32\textwidth]{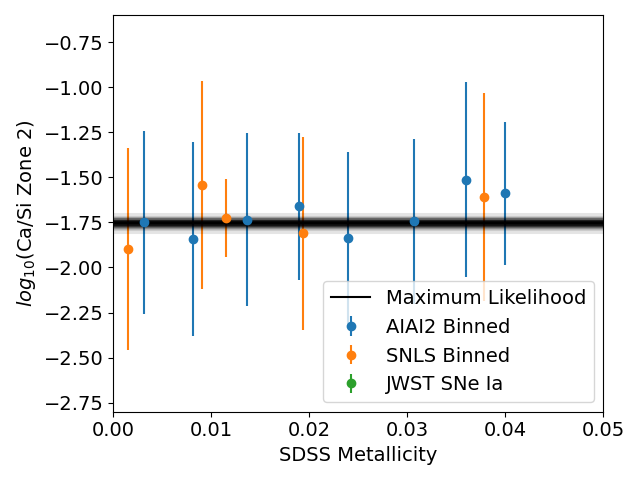}
    \includegraphics[width=0.32\textwidth]{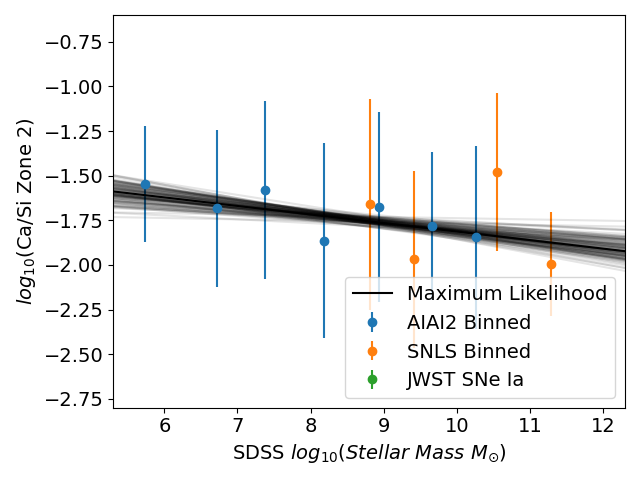}
    \caption{The correlation between the Ca/Si elemental abundances and host galaxy mean age (left column), metallicity (middle column), and stellar mass (right column). From up to down, the first and the third row use the host galaxy properties from DESI, the second and the fourth row use the host galaxy properties from SDSS, the first and the second row uses the Ca/Si ratio in Zone 2, the third and the fourth row uses the Ca/Si ratio in Zone 3. Note that in the first and the third figure of the right column, the host galaxy stellar mass data of SN~2025ogs and SN~H0pe are adopted from \cite{Frye2024SNH0pe,Siebert2026SN2025ogs}. }\label{fig:hostgalaxy}
\end{figure*}

In Figure \ref{fig:hostgalaxy}, we compare the Ca/Si abundance ratios in Zones 2 and 3 with host-galaxy stellar mass, metallicity, and mean stellar age from the DESI and SDSS surveys. 
For illustration purpose, we bin the SNe Ia data from \citetalias{Chen2024AIAI2} and the SNLS project into 10 uniform bins. 
We find a positive correlation between the Zone 3 Ca/Si ratio and DESI host-galaxy metallicity, and a negative correlation between the Zone 2 Ca/Si ratio and DESI host-galaxy stellar age. However, neither correlation is reproduced in the SDSS data. 
A weak negative correlation between the host galaxy stellar mass and the Ca/Si ratio in either Zone 2 or Zone 3 is observed in both the DESI and SDSS measurements. 

\subsection{Error Estimate}

\begin{figure*}
    \centering
    \plottwo{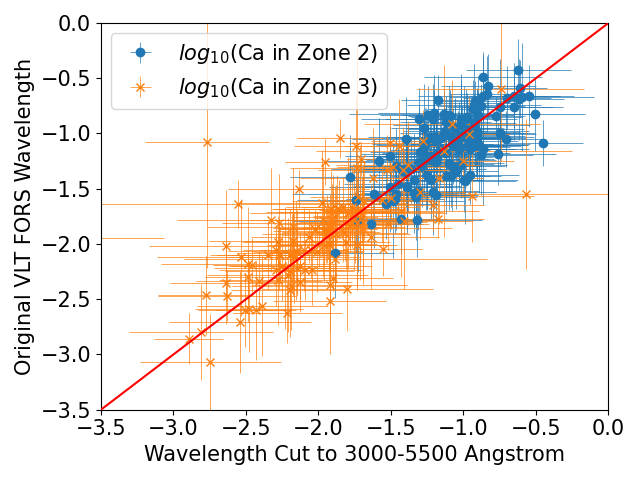}{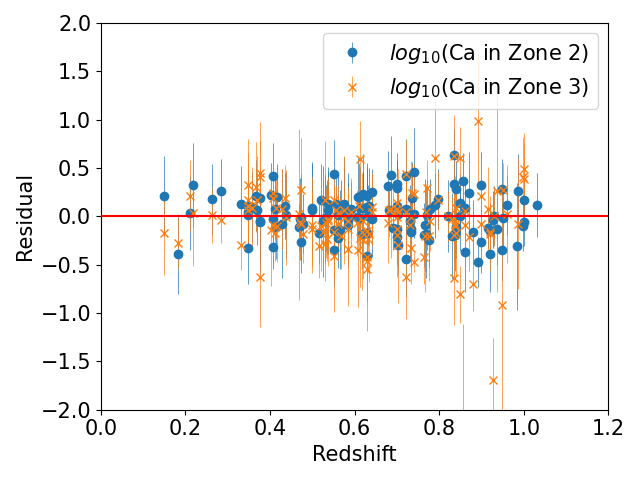}
    \caption{Left Panel: The Ca elemental abundances in Zone 2 and 3 estimated by the AIAI method, x axis values use the spectra with a wavelength cut to 3000-5500 $\rm\AA$, while the y axis values use the original wavelength range of VLT. Right Panel: The residual of the Ca elemental abundances, with and without wavelength cut, as a function of redshift. }
    \label{fig:CaError}
\end{figure*}

A large fraction of the spectra used in this work were obtained with ground-based optical telescopes and are limited to observer-frame wavelengths between 3000 and 10,000 $\rm\AA$. 
As SN Ia redshift increases, the observable rest-frame wavelengths are systematically blueshifted. This shift can include or exclude particular spectral lines and may introduce systematic bias in elemental-abundance estimates. 
Although the AIAI deep-learning uncertainties account for wavelength-coverage effects, we further assess the elemental-abundance accuracy and possible systematic bias using the 139 VLT spectra. We restrict the wavelength range to 3000--5500 $\rm\AA$ for AIAI predictions and compare the results with those obtained using the full wavelength range, as shown in Figure \ref{fig:CaError}. 
The Ca abundance predictions with the restricted wavelength range are consistent with those obtained using the full wavelength range, and we find no redshift-dependent prediction bias. 
Therefore, we conclude that the wavelength restriction does not significantly affect the Ca abundance estimates. 

In Appendix \ref{sec:microlensing}, we discuss the possible impact of the microlensing effect on the elemental abundance estimation results. 

\section{Discussion}\label{sec:discussion}

We use the deep-learning SNe Ia ejecta estimator from \citetalias{Chen2024AIAI2} to estimate the elemental abundances of 201 SNe Ia observed by SNLS and two lensed SNe Ia, extending the sample to redshift 2.05. 
We find that high-redshift SNe Ia contain more Ca in their ejecta. 
From redshift 0 to redshift 2, the Ca/Si ratio increases by $0.79\pm0.12$ dex in Zone 2 and $0.50\pm0.14$ dex in Zone 3. 
In comparison, nucleosynthesis simulations indicate that, for a fixed explosion hydrodynamic process, variation in progenitor metallicity can account for only $0.1\sim0.2$ dex of the Ca/Si-ratio variation. 
The discrepancy between the observed and simulated evolution of the Ca/Si ratio suggests that the high-redshift SN Ia population may arise from different progenitor systems or explosion hydrodynamic processes, than those in low redshift. 

Moreover, \cite{Balland2009SNLSVLT} split the SNLS SNe Ia spectra observed by VLT into low redshift bin ($z<0.5$) and high redshift bin ($z\leq 0.5$) and discovered that the average spectra in the high redshift bin show stronger Ca II H\&K spectral line than the average spectra in the low redshift bin, which is qualitatively consistent with our discoveries. 

Cross-matching the SNe Ia host galaxies with the DESI and SDSS spectral survey results, we observe weak negative correlation between the SNe Ia Ca/Si ratio and the host galaxy stellar mass, 
Further validation on this correlation, will rely on systematic surveys on both high-redshift SNe Ia and host galaxies, to accumulate an unbiased sample set. 
Alternatively, more detailed stellar polulation analyses; comparative studies between the SNe Ia host galaxies and low-SN-rate galaxies; which have proven useful on SNe Ia host galaxies' integral field unit spectra (e.g., \citet{Chen2021SFH}), could probe the environmental dependence of SNe Ia explosions, in order to identify the low-redshift counterparts of the high-redshift, high Ca/Si ratio SNe Ia host galaxies. 


The most remote SN Ia that have been spectroscopically observed is SN~2023adsy at redshift $z\approx 2.9$ \citep{Pierel2024SN2023adsy}, while several other SNe Ia at $z\leq 1.5$ (e.g. SN Encore, \cite{Pierel2024SNEncore}) are observed. 
However, these spectra are not included in this study because the current AIAI program only covers the SNe Ia with phases between $-10$ days and $20$ days, and simulating the late-phase SNe Ia requires more detailed non-local thermodynamic equilibrium radiative transfer calculations. 

\begin{acknowledgments}

X.C. is supported by Texas A\&M University Institute of Data Science (TAMIDS). 
The work of U. B-N. is supported by the National Science Foundation under award CCF-2225507. 
Portions of this research were conducted with the advanced computing resources provided by Texas A\&M High Performance Research Computing. 
The authors would like to thank Dr. Ken Shen from UC Berkeley, Prof. Dean Townsley from University of Alabama for supportive discussions. 

\end{acknowledgments}

\software{SEDONA\citep{Kasen2006Sedona}, TARDIS\citep{Kerzendorf2014Tardis}}

\bibliography{sample631cita}{}
\bibliographystyle{aasjournal}

\appendix

\section{The SNe Ia List}\label{sec:sntable}

In addition to the 124 SNe Ia listed in \citetalias{Chen2024AIAI2}, this work uses SNe Ia from the SNLS project and two SNe Ia observed by JWST. 
Table \ref{tab:Lensed} lists the SNe Ia observed by JWST and selected in this research. 
Table \ref{tab:SNLSKeck} lists the SNe Ia observed by Keck telescope through the SNLS project as described in \cite{Ellis2008SNLSKeck}. 
Table \ref{tab:SNLSVLT} lists the SNe Ia observed by VLT through the SNLS project as described in \cite{Balland2009SNLSVLT}. 
Table \ref{tab:SNLSGemini} lists the SNe Ia observed by Gemini North or South telescope through the SNLS project as described in \cite{Howell2005Gemini}. 
Table \ref{tab:SNLSGemini2} lists the SNe Ia observed by Gemini North or South telescope through the SNLS project as described in \cite{Bronder2008Gemini}. 

\begin{table}[htb!]
    \centering
    \begin{tabular}{ccccc}
        \hline
        Supernova Name & Redshift & Telescope & Phase & MSE \\
        \hline
        \hline
        SN 2025ogs & 2.05 & JWST      & 1.0 & 0.105 \\
        SN H0pe    & 1.78 & JWST      & 4.2 & 0.069 \\
        \hline
    \end{tabular}
    \caption{The high-redshift SNe Ia selected in this research. }
    \label{tab:Lensed}
\end{table}

\begin{table}[htb!]
    \centering
    \begin{tabular}{cccc|cccc|cccc}
        \hline
        SN Name & Redshift & Phase & MSE & SN Name & Redshift & Phase & MSE & SN Name & Redshift & Phase & MSE \\
        \hline
        \hline
        03D1au & 0.5043 & -1.6 & 0.094 & 03D1aw & 0.582 & 2.4 & 0.129 & 03D1co & 0.679 & 7.1 & ... \\ 
        03D1dj & 0.4 & -1.8 & 0.04 & 03D3af & 0.532 & 2.8 & 0.119 & 03D3aw & 0.449 & -0.8 & 0.126 \\ 
        03D3ay & 0.3709 & -1.0 & 0.042 & 03D3ba & 0.2912 & 12.8 & 0.143 & 03D3bb & 0.2437 & 2.2 & 0.061 \\ 
        03D3bh & 0.2486 & -3.5 & 0.033 & 03D3bl & 0.3553 & 3.7 & 0.139 & 03D3cc & 0.4627 & 8.5 & 0.108 \\ 
        03D3cd & 0.4607 & -5.6 & 0.046 & 03D4ag & 0.2847 & -5.0 & 0.019 & 03D4cj & 0.27 & -7.4 & 0.089 \\ 
        03D4dh & 0.6268 & 0.5 & 0.124 & 04D1hd & 0.3688 & 1.6 & 0.022 & 04D1jg & 0.5842 & -2.0 & 0.078 \\ 
        04D1oh & 0.59 & -6.0 & 0.509 & 04D1rh & 0.4349 & 2.6 & 0.073 & 04D1sk & 0.6634 & -0.4 & 0.173 \\ 
        04D2gc & 0.5216 & -0.2 & 0.079 & 04D2kr & 0.7441 & -2.5 & 0.185 & 04D3cp & 0.83 & 5.2 & 0.339 \\ 
        04D3ez & 0.263 & 1.6 & 0.027 & 04D3fk & 0.3578 & -6.9 & 0.047 & 04D4in & 0.516 & -4.6 & 0.048 \\ 
        04D4jr & 0.482 & -0.2 & 0.046 & 05D1hk & 0.2631 & -8.6 & 0.038 & 05D1hn & 0.1489 & -7.2 & 0.042 \\ 
        05D1if & 0.763 & -5.0 & 0.12 & 05D1ix & 0.49 & -8.7 & 0.093 & 05D1iy & 0.2478 & -9.2 & 0.022 \\ 
        05D2le & 0.7002 & 6.3 & 0.209 & 05D2mp & 0.3537 & -3.9 & 0.045 & \\ 
        \hline
    \end{tabular}
    \caption{The list of SNe Ia spectra observed by Keck, reported in \cite{Ellis2008SNLSKeck}. The MSE values are derived from Equation~\ref{eq:mse}. Note that the AIAI fitting methods failed on 1 spectra and no MSE values are available. }
    \label{tab:SNLSKeck}
\end{table}

\begin{table}
    \centering
    \begin{tabular}{cccc|cccc|cccc}
        \hline
        SN Name & Redshift & Phase & MSE & SN Name & Redshift & Phase & MSE & SN Name & Redshift & Phase & MSE \\
        \hline
        \hline
        05D2fq & 0.733 & 2.07 & 0.412 & 04D1ow & 0.915 & 6.369 & 0.247 & 04D4jw & 0.961 & 2.21 & 0.29 \\ 
        05D4ef & 0.605 & 3.46 & 0.287 & 04D2an & 0.62 & -3.39 & 0.068 & 04D2an & 0.62 & 0.32 & 0.069 \\ 
        06D1ab & 0.182 & -4.41 & 0.063 & 04D1jd & 0.778 & 6.816 & 0.304 & 06D2bk & 0.499 & 0.86 & 0.104 \\ 
        04D4ht & 0.217 & 6.57 & 0.18 & 05D4bj & 0.701 & -1.95 & 0.171 & 04D2ac & 0.348 & 1.31 & 0.069 \\ 
        05D4dw & 0.855 & 4.68 & 0.856 & 06D4cl & 1.0 & -2.5 & ... & 05D2ci & 0.63 & 3.63 & ... \\ 
        05D2ci & 0.63 & 4.87 & ... & 05D2ci & 0.63 & 6.72 & ... & 05D2bw & 0.92 & 1.94 & 0.438 \\ 
        04D4ib & 0.699 & 0.66 & 0.104 & 04D1pp & 0.735 & 2.208 & 0.369 & 04D4an & 0.613 & 8.07 & 0.593 \\ 
        04D1sa & 0.585 & -2.572 & 0.158 & 03D4gg & 0.592 & 16.42 & 0.397 & 05D4af & 0.499 & 11.31 & 0.133 \\ 
        05D4cq & 0.701 & -0.14 & 0.093 & 04D1ak & 0.526 & 11.81 & 0.623 & 04D2mc & 0.348 & 6.46 & 0.103 \\ 
        05D2ct & 0.734 & 7.82 & 0.241 & 05D4cs & 0.79 & -1.7 & 0.093 & 04D4hf & 0.936 & -0.52 & 0.498 \\ 
        06D2cb & 1.0 & 2.99 & ... & 05D2ay & 0.92 & 5.78 & ... & 05D1iz & 0.86 & 7.8 & 0.223 \\ 
        05D1ke & 0.69 & 2.16 & 0.086 & 05D2by & 0.891 & 0.91 & ... & 05D2nn & 0.87 & -1.25 & ... \\ 
        04D4ju & 0.472 & -2.18 & 0.111 & 03D1gt & 0.56 & 6.997 & 0.239 & 05D2eb & 0.534 & -4.73 & 0.066 \\ 
        04D2fs & 0.357 & 1.73 & 0.029 & 04D1hx & 0.56 & 5.939 & 0.344 & 05D4be & 0.537 & 3.88 & 0.079 \\ 
        05D1cl & 0.83 & 8.73 & ... & 03D4dy & 0.61 & 4.77 & 0.077 & 04D1qd & 0.767 & -0.158 & 0.103 \\ 
        03D4cx & 0.949 & 1.15 & ... & 03D4cx & 0.949 & 2.69 & 0.25 & 04D2cf & 0.369 & 8.48 & 0.144 \\ 
        04D1kj & 0.585 & -3.72 & 0.064 & 04D2cc & 0.838 & 6.07 & ... & 04D2cc & 0.838 & 7.16 & ... \\ 
        04D1rx & 0.985 & 0.91 & ... & 03D1ar & 0.408 & 5.305 & 0.245 & 04D1ks & 0.798 & -1.011 & 0.194 \\ 
        04D1ag & 0.557 & 4.303 & 0.076 & 06D2ce & 0.82 & 0.0 & 0.055 & 05D2bv & 0.474 & -0.12 & 0.052 \\ 
        04D1pd & 0.95 & 2.469 & 0.222 & 05D2dw & 0.417 & -5.25 & 0.038 & 03D1fl & 0.687 & 0.525 & 0.127 \\ 
        06D2ca & 0.531 & 0.0 & 0.16 & 06D2ca & 0.531 & 0.65 & 0.208 & 04D2ja & 0.74 & 9.18 & 0.891 \\ 
        04D2ja & 0.74 & 9.76 & 1.17 & 04D1iv & 0.998 & 3.01 & ... & 06D4co & 0.437 & 3.48 & 0.052 \\ 
        04D2al & 0.836 & -2.48 & 0.25 & 04D4id & 0.769 & 2.95 & 0.16 & 05D4cn & 0.763 & 4.65 & 0.256 \\ 
        03D4di & 0.899 & -8.64 & ... & 03D4di & 0.899 & -6.54 & ... & 03D1fc & 0.332 & -4.386 & 0.029 \\ 
        04D2gp & 0.732 & 2.73 & 0.189 & 05D2ie & 0.348 & -8.9 & 0.062 & 03D1bm & 0.575 & -5.135 & 0.192 \\ 
        03D4cy & 0.927 & 4.57 & 0.71 & 04D4jr & 0.47 & -5.93 & 0.039 & 05D4ag & 0.64 & 14.03 & 0.409 \\ 
        03D4au & 0.468 & 6.47 & 0.404 & 05D4ay & 0.408 & 8.16 & 0.284 & 05D4ay & 0.408 & 9.58 & 0.333 \\ 
        03D4ag & 0.285 & -8.64 & 0.013 & 04D2gc & 0.521 & -4.94 & 0.266 & 04D4gz & 0.375 & -5.82 & 0.155 \\ 
        04D4gz & 0.375 & 10.18 & 0.365 & 06D2ga & 0.84 & 5.43 & 0.297 & 05D4ek & 0.536 & 2.08 & 0.047 \\ 
        06D4ce & 0.85 & 2.72 & ... & 06D4ce & 0.85 & 3.26 & ... & 03D1dt & 0.612 & 5.07 & 0.239 \\ 
        05D2cb & 0.427 & -5.78 & 0.149 & 05D1hk & 0.263 & -4.96 & 0.058 & 05D1ck & 0.617 & -2.65 & 0.949 \\ 
        05D4ej & 0.585 & 7.58 & 0.113 & 05D4bi & 0.775 & -1.51 & 0.091 & 05D2he & 0.608 & 2.99 & 0.185 \\ 
        04D1aj & 0.721 & 11.63 & 0.348 & 04D1aj & 0.721 & 13.38 & 0.432 & 04D4dw & 1.031 & 2.1 & ... \\ 
        03D1co & 0.679 & -4.127 & 0.143 & 04D2iu & 0.7 & 9.81 & 0.641 & 04D2iu & 0.7 & 10.4 & 0.524 \\ 
        05D2ec & 0.64 & 2.66 & 0.21 & 04D2ca & 0.835 & 12.01 & ... & 04D2ca & 0.835 & 13.11 & ... \\ 
        04D1pg & 0.515 & -1.275 & 0.055 & 05D1cb & 0.632 & 4.25 & 0.145 & 03D1bf & 0.703 & -2.681 & 0.262 \\ 
        04D1pc & 0.77 & 0.07 & 0.169 & 03D4at & 0.634 & 5.48 & 0.282 & 05D4fg & 0.839 & -0.26 & 0.159 \\ 
        04D1ff & 0.86 & 4.62 & 0.214 & 04D1si & 0.702 & -1.718 & 0.088 & 05D4ev & 0.722 & -3.61 & 0.121 \\ 
        04D1dc & 0.211 & -0.409 & 0.021 & 05D1dn & 0.566 & -4.7 & 0.076 & 06D4cq & 0.411 & -1.42 & 0.061 \\ 
        05D4ff & 0.402 & 5.11 & 0.801 & 05D4cw & 0.375 & 6.95 & 0.064 & 04D4bk & 0.88 & 3.14 & ... \\ 
        05D2ei & 0.366 & 16.86 & 0.219 & 06D2ck & 0.552 & 7.72 & 0.227 & 03D1bp & 0.347 & -7.5 & 0.129 \\ 
        05D4fe & 0.984 & -1.95 & ... & 04D2cw & 0.568 & 19.28 & 0.414 & 06D2cc & 0.532 & 3.26 & 0.09 \\ 
        06D2cd & 0.93 & 4.15 & 0.091 & 05D2dt & 0.574 & -1.694 & 0.076 & 05D1hn & 0.149 & -1.03 & 0.063 \\ 
        04D4bq & 0.55 & 2.55 & 0.186 & 04D4bq & 0.55 & 5.13 & 0.121 & 04D4fx & 0.629 & -8.35 & 0.111 \\ 
        05D2ac & 0.479 & 2.04 & 0.048 & 04D2fp & 0.415 & 1.81 & 0.033 & 05D2bt & 0.68 & 1.4 & 0.192 \\ 
        04D1rh & 0.436 & 0.047 & 0.057 & \\
        \hline
    \end{tabular}
    \caption{The list of SNe Ia spectra observed by VLT during the SNLS, reported in \cite{Balland2009SNLSVLT}. The MSE values are derived from Equation~\ref{eq:mse}. Note that the AIAI fitting methods failed on 23 spectra and no MSE values are available. }
    \label{tab:SNLSVLT}
\end{table}

\begin{table}
    \centering
    \begin{tabular}{cccc|cccc|cccc}
        \hline
        SN Name & Redshift & Phase & MSE & SN Name & Redshift & Phase & MSE & SN Name & Redshift & Phase & MSE \\
        \hline
        \hline
        03D1ax & 0.496 & -3.3 & 0.051 & 03D1bk & 0.865 & -2.1 & 0.215 & 03D1cm & 0.87 & 2.4 & 1.041 \\ 
        03D1co & 0.68 & -1.2 & 1.668 & 03D1fq & 0.8 & 2.2 & 0.444 & 03D4cj & 0.27 & -6.9 & 0.231 \\ 
        03D4cn & 0.818 & 2.8 & 0.437 & 03D4cz & 0.695 & 8.7 & 0.251 & 03D4fd & 0.791 & -1.5 & 0.781 \\ 
        03D4gl & 0.56 & -6.0 & 1.037 & 04D1hd & 0.3685 & -4.0 & 0.187 & 04D1hy & 0.85 & -3.2 & 0.432 \\ 
        04D1ow & 0.93 & -0.8 & 0.759 & 04D3bf & 0.156 & 15.1 & 0.066 & 04D3dd & 1.01 & -1.4 & 0.803 \\ 
        04D3fq & 0.73 & 3.6 & 0.738 & 04D3hn & 0.5516 & 4.9 & 0.175 & 04D3kr & 0.3373 & 1.7 & 0.117 \\ 
        04D3lu & 0.8218 & 5.6 & 0.269 & 04D3mk & 0.813 & 1.2 & 0.241 & 04D3ml & 0.95 & 1.1 & 0.456 \\ 
        04D3nh & 0.3402 & 3.7 & 0.107 & 04D3nq & 0.22 & 8.8 & 0.083 & 04D3ny & 0.81 & 2.5 & 0.364 \\ 
        04D3oe & 0.756 & 1.8 & 0.228 & 04D4dm & 0.811 & -0.7 & 0.73 & 04D4gg & 0.4238 & -10.0 & 0.049 \\ 
        04D4hu & 0.7027 & 5.0 & 0.154 & 04D4ic & 0.68 & 4.8 & 0.421 & 04D4ii & 0.866 & -3.5 & 0.289 \\ 
        04D4im & 0.751 & 0.2 & 0.181 & \\
        \hline
    \end{tabular}
    \caption{The list of SNe Ia spectra observed by Gemini, the phase data is reported in \cite{Howell2005Gemini} using the spectral fitting method. The MSE values are derived from Equation \ref{eq:mse}. }
    \label{tab:SNLSGemini}
\end{table}

\begin{table}[htb!]
    \centering
    \begin{tabular}{cccc|cccc|cccc}
        \hline
        SN Name & Redshift & Phase & MSE & SN Name & Redshift & Phase & MSE & SN Name & Redshift & Phase & MSE \\
        \hline
        \hline
        03D1ax & 0.5 & -2.3 & 0.043 & 03D1bk & 0.87 & -5.23 & 0.231 & 03D1cm & 0.87 & -4.42 & 1.055 \\ 
        03D1co & 0.679 & 7.42 & 1.944 & 03D4cj & 0.27 & -8.05 & 0.225 & 03D4cn & 0.82 & 0.53 & 0.464 \\ 
        03D4fd & 0.79 & -0.88 & 0.744 & 03D4gl & 0.56 & 8.27 & 1.002 & 04D1de & 0.77 & -6.8 & 0.095 \\ 
        04D1hd & 0.37 & 1.85 & 0.189 & 04D1hy & 0.85 & -2.19 & 0.432 & 04D1ow & 0.915 & 6.28 & 0.751 \\ 
        04D2mh & 0.59 & 1.58 & 0.287 & 04D3dd & 1.01 & 3.68 & 0.808 & 04D3fq & 0.73 & 1.35 & 0.629 \\ 
        04D3kr & 0.34 & 5.37 & 0.131 & 04D3lp & 0.98 & 1.0 & 0.741 & 04D3mk & 0.81 & -1.47 & 0.234 \\ 
        04D3ml & 0.95 & -0.89 & 0.507 & 04D3nq & 0.22 & 9.36 & 0.09 & 04D3ny & 0.81 & 2.15 & 0.375 \\ 
        04D4dm & 0.81 & 3.55 & 0.579 & 04D4hu & 0.7 & 6.22 & 0.148 & 04D4ic & 0.68 & 2.92 & 0.423 \\ 
        04D4ii & 0.87 & -4.58 & 0.281 & 05D1az & 0.84 & 8.85 & 0.545 & 05D1cc & 0.56 & 6.8 & 0.163 \\ 
        05D1er & 0.85 & 2.83 & 0.352 & 05D1ju & 0.71 & 6.28 & 0.284 & 05D1kl & 0.56 & -4.27 & 0.152 \\ 
        05D2ab & 0.32 & -1.3 & 0.077 & 05D2ah & 0.18 & -2.49 & 0.069 & 05D2ja & 0.3 & 9.63 & 0.079 \\ 
        05D3ax & 0.64 & 8.03 & 0.147 & 05D3cf & 0.42 & 13.68 & 0.136 & 05D3cq & 0.89 & 11.6 & 0.753 \\ 
        05D3cx & 0.81 & 10.28 & 0.565 & 05D3km & 0.97 & 1.46 & 1.351 & 05D3kt & 0.65 & -3.86 & 0.158 \\ 
        05D3lb & 0.65 & -1.21 & 0.15 & 05D3mh & 0.67 & 2.71 & 0.416 & 05D3mn & 0.76 & 2.82 & 0.61 \\ 
        05D3mq & 0.25 & 8.45 & 0.142 & 05D4av & 0.51 & 5.84 & 0.181 & 05D4bm & 0.38 & -3.81 & 0.05 \\ 
        05D4cn & 0.763 & 0.39 & 0.141 & 05D4dy & 0.79 & 2.13 & 0.57 & 05D4gw & 0.81 & 0.13 & 0.355 \\ 
        \hline
    \end{tabular}
    \caption{The list of SNe Ia spectra observed by Gemini, reported in \cite{Bronder2008Gemini}. The MSE values are derived from Equation \ref{eq:mse}.  }
    \label{tab:SNLSGemini2}
\end{table}

\clearpage

\section{The SNLS SNe Ia Spectra}\label{sec:snspectra}

\begin{figure}[htb!]
    \includegraphics[width=0.5\textwidth]{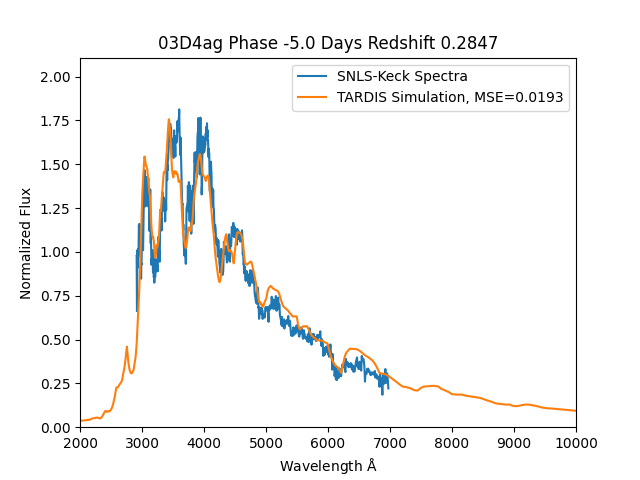}
    \includegraphics[width=0.5\textwidth]{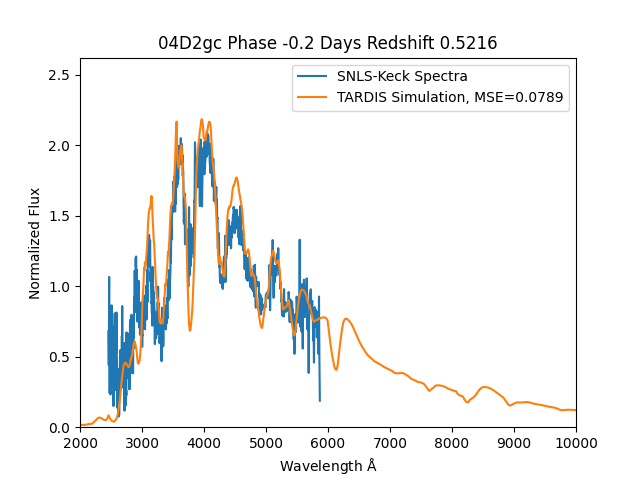}
    \includegraphics[width=0.5\textwidth]{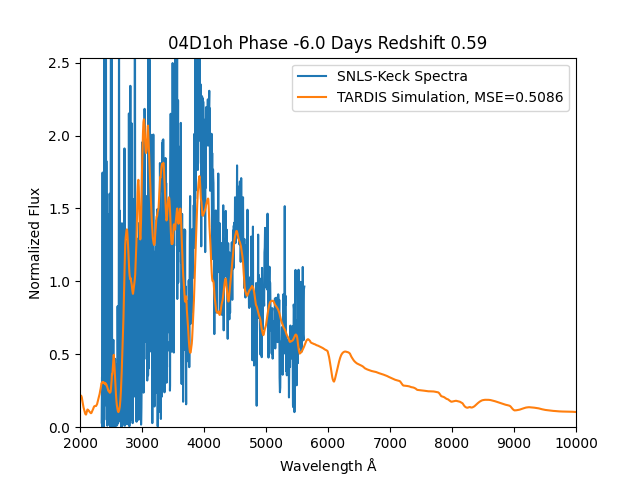}
    \includegraphics[width=0.5\textwidth]{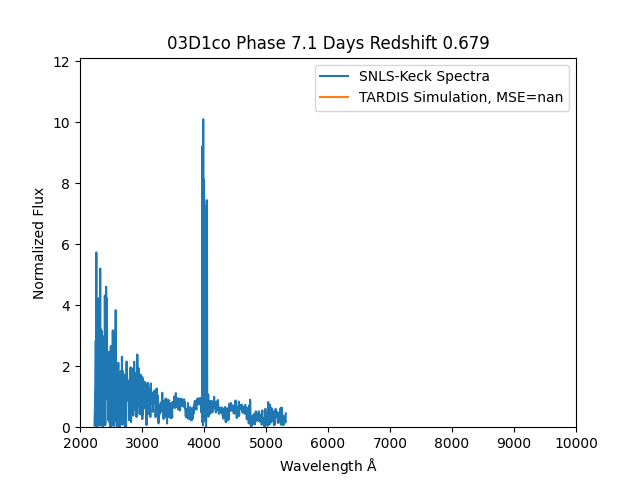}
    \caption{Selected spectra of SNe Ia observed by Keck (blue line), and the TARDIS simulation spectra (orange line). The upper left panel shows the spectra of SNLS-03D4ag, which has the smallest spectral fitting MSE values among all the Keck observed spectra; the upper right panel shows the spectra of SNLS-04D2gc, which has the median MSE value; the lower left panel shows the spectra of SNLS-04D1oh, which has the largest MSE value; the lower right panel shows the spectra of SNLS-03D1co, which is failed in AIAI method possibly due to the large spectral noise around rest frame 4000 $\rm\AA$}\label{fig:SNLSKeck}
\end{figure}

\begin{figure}
    \includegraphics[width=0.33\textwidth]{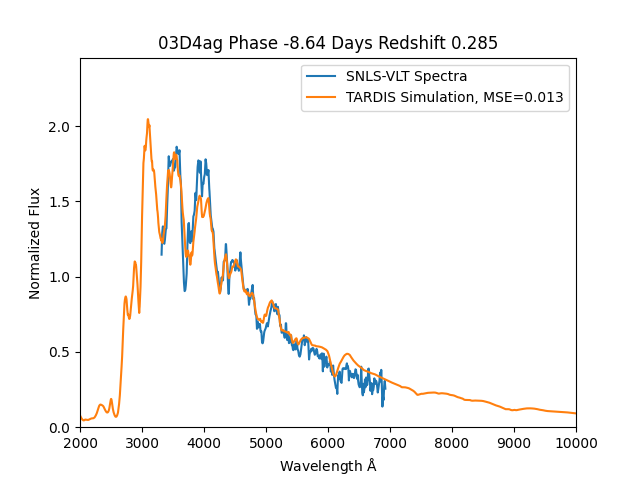}
    \includegraphics[width=0.33\textwidth]{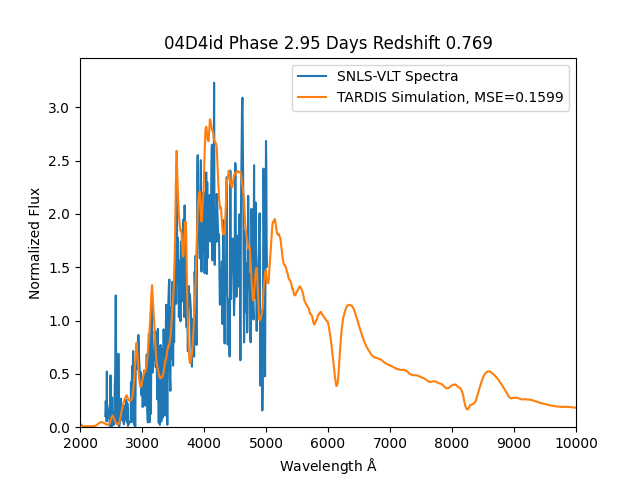}
    \includegraphics[width=0.33\textwidth]{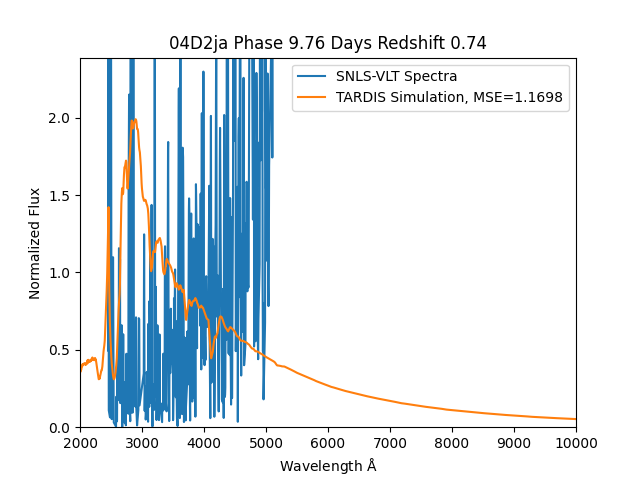}
    \caption{Selected spectra of SNe Ia observed by VLT (blue line), and the TARDIS simulation spectra (orange line). The left panel shows the spectra of SNLS-03D4ag, which has the smallest spectral fitting MSE values among all the VLT observed spectra; the middle panel shows the spectra of SNLS-04D4id, which has the median MSE value; the right panel shows the spectra of SNLS-04D2ja, which has the largest MSE value. }\label{fig:SNLSVLT}
\end{figure}

\begin{figure}
    \includegraphics[width=0.33\textwidth]{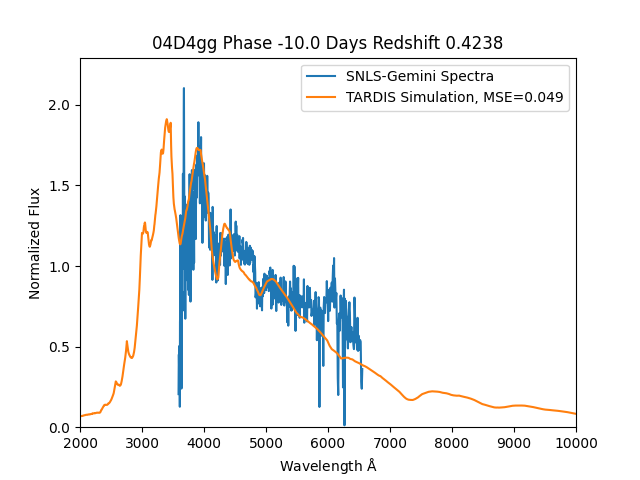}
    \includegraphics[width=0.33\textwidth]{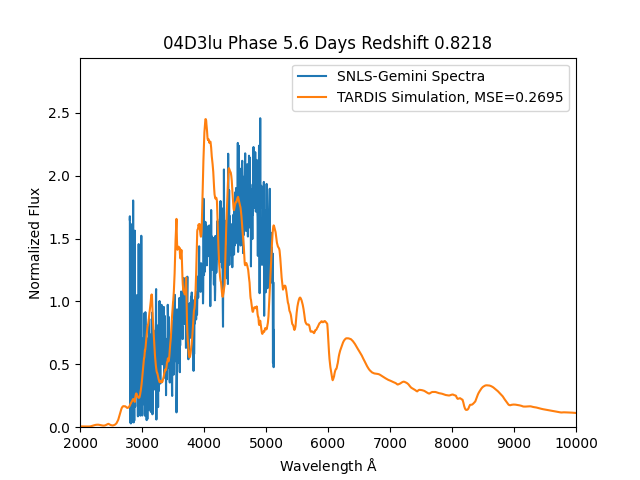}
    \includegraphics[width=0.33\textwidth]{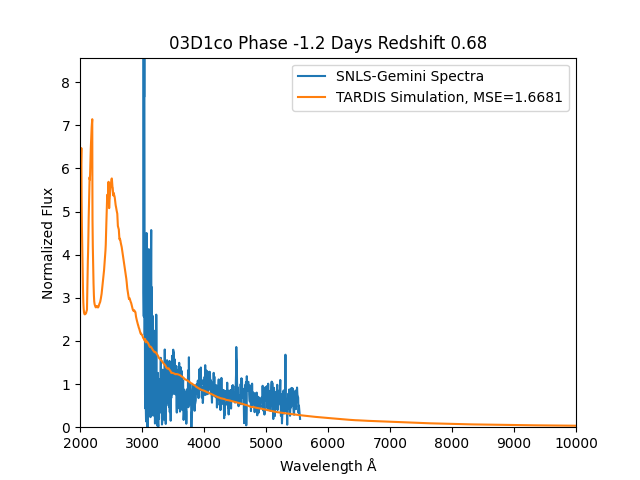}
    \caption{Selected spectra of SNe Ia observed by Gemini (blue line) and reported in \cite{Howell2005Gemini}, and the TARDIS simulation spectra (orange line). The left panel shows the spectra of SNLS-04D4gg, which has the smallest spectral fitting MSE values among all the Gemini observed spectra; the middle panel shows the spectra of SNLS-04D3lu, which has the median MSE value; the right panel shows the spectra of SNLS-03D1co, which has the largest MSE value. }\label{fig:SNLSGemini}
\end{figure}

\begin{figure}
    \includegraphics[width=0.33\textwidth]{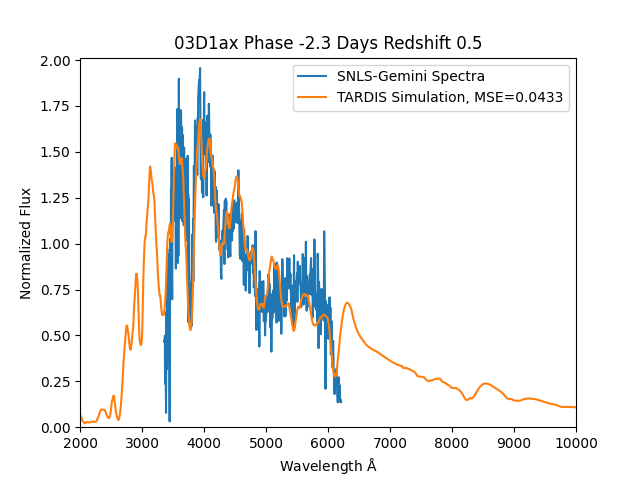}
    \includegraphics[width=0.33\textwidth]{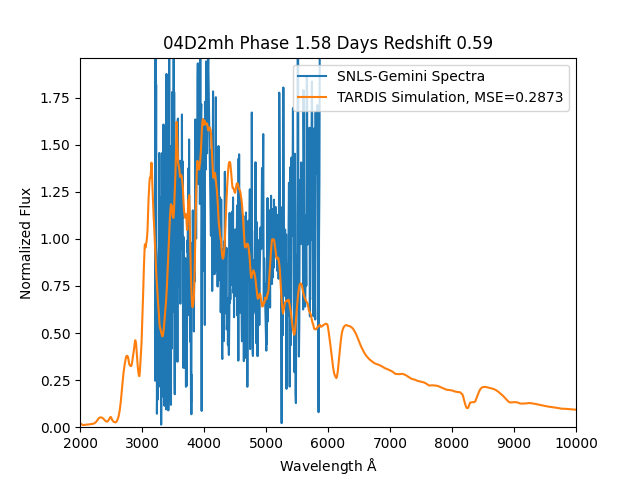}
    \includegraphics[width=0.33\textwidth]{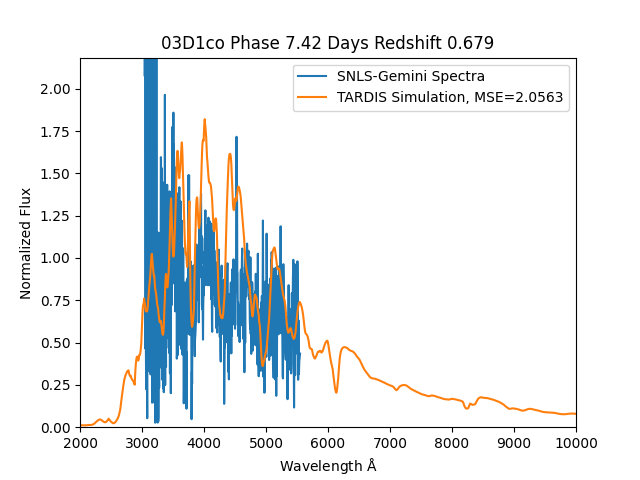}
    \caption{Selected spectra of SNe Ia observed by Gemini (blue line) and reported in \cite{Bronder2008Gemini}, and the TARDIS simulation spectra (orange line). The left panel shows the spectra of SNLS-04D4gg, which has the smallest spectral fitting MSE values among all the Gemini observed spectra; the middle panel shows the spectra of SNLS-04D3lu, which has the median MSE value; the right panel shows the spectra of SNLS-03D1co, which has the largest MSE value. }\label{fig:SNLSGemini2}
\end{figure}

This section shows selected observed SN Ia spectra and the corresponding TARDIS radiative-transfer simulations. 
Figure \ref{fig:SNLSKeck} shows SN Ia spectra observed with Keck through SNLS, as described by \citet{Ellis2008SNLSKeck}. 
Figure \ref{fig:SNLSVLT} shows SN Ia spectra observed with the VLT through SNLS, as described by \citet{Balland2009SNLSVLT}. 
Figure \ref{fig:SNLSGemini} shows SN Ia spectra observed with Gemini North or South through SNLS, as described by \citet{Howell2005Gemini}. 
Figure \ref{fig:SNLSGemini2} shows SN Ia spectra observed with Gemini North or South through SNLS, as described by \citet{Bronder2008Gemini}. 
Only spectra with the minimum, median, and maximum MSE values calculated using Equation \ref{eq:mse} are shown; the remaining spectra will be provided in the Zenodo online material \footnote{\href{https://zenodo.org/uploads/21877558}{https://zenodo.org/uploads/21877558}}. 

\clearpage

\section{Microlensing}\label{sec:microlensing}

Radiative-transfer simulations have predicted chromatic microlensing effects in SNe Ia beginning at $\sim$20 days after explosion (e.g., \citealt{Goldstein2018Micrloensing}). 
To quantitatively estimate the effects of microlensing on SN Ia elemental-abundance estimates, we use two microlensing magnification maps from the GERLUMPH dataset \citep{Vernardos2014Gerlumph,Vernardos2015Gerlumph} to simulate microlensed SN Ia spectra. We then use the deep-learning estimator from \citetalias{Chen2024AIAI2} to predict elemental mass fractions from these spectra. 
The spatially-resolved SN Ia data cube is simulated on the three-dimensional N100 deflagration-to-detonation model \citep{Seitenzahl2013N100} using the SEDONA-GesaRaT radiative transfer program \citep{Chen2025GesaRaT}. 
The SN Ia and gravitational-lens redshifts are set to 1.78 and 0.348, respectively, matching SN~H0pe. 

In the GERLUMPH dataset, microlensing magnification maps are generated using stellar objects with masses of $\langle M\rangle=1M_\odot$ randomly positioned on a square field with a width of 25 Einstein radii ($R_{\rm ein}$), where $R_{\rm ein}$ is defined as

\begin{equation}
    R_{ein}=\sqrt{\frac{D_{os}D_{ls}}{D_{ol}}\frac{4G \langle M \rangle}{c^2}} \ , 
\end{equation}
where $D_{os}$, $D_{ls}$, and $D_{ol}$ are the angular-diameter distances between the observer and source, lens and source, and observer and lens, respectively, and $G$ is the gravitational constant. 
The gravitational-lens mass distribution is controlled by three parameters: the convergence parameter $\kappa$, which controls the combined focusing power of stellar objects and smooth matter; the shear parameter $\gamma$, which controls distortion due to the external mass distribution outside the field; and the smooth-mass fraction $s$, which controls the ratio of stellar objects to smooth matter. 
In this study, the choices of $\kappa$ and $\gamma$, as shown in Table \ref{tab:lensparameter}, are broadly consistent to the SN H0pe measurements on 3 lensing images \citep{Pierel2024SNH0pe}. 
The parameters ($s$) of SN H0pe are assumed to be 0.99 because the galaxy cluster gravitational lens is dominated by dark matter with a smooth mass profile, while we select $s=0$ in order to distinguish the chromatic distortion on SNe Ia spectra due to microlensing effect. 

\begin{figure*}[htb!]
    \centering
    \plottwo{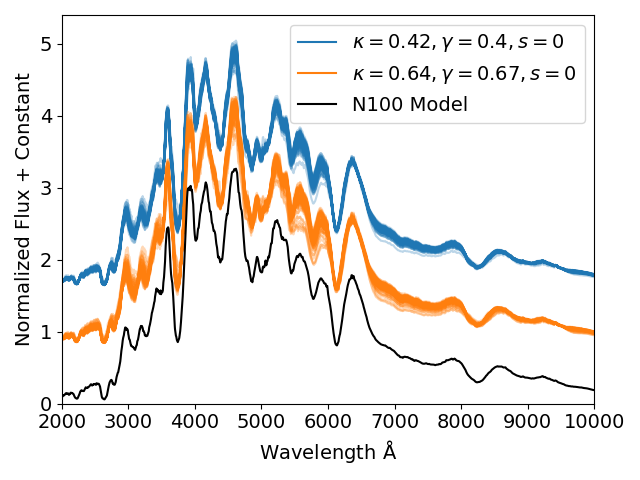}{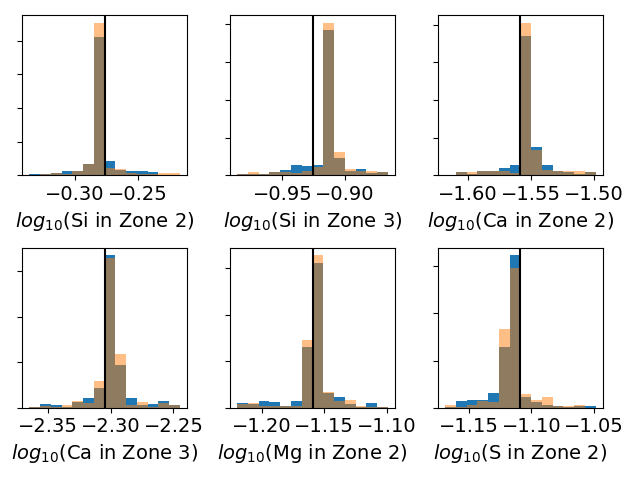}
    \caption{Left Panel: The simulated spectra of the N100 explosion model at 20 days after the explosion. Right panel: The distribution histogram of the AIAI predicted elemental abundances using the microlensed spectra, compared to the predictions using the unlensed spectra. The blue lines in the left panel and the blue histogram in the right panel uses the first microlensing map listed in Table \ref{tab:lensparameter}, the orange lines in the left panel and the orange transparent histogram uses the second microlensing map listed in Table \ref{tab:lensparameter}, the black line is the result without microlensing. }\label{fig:microlens}
\end{figure*}

In Figure \ref{fig:microlens}, we randomly select 300 SN coordinates in each magnification map, calculate normalized N100 SN Ia spectra with microlensing at 20 days after explosion, and apply AIAI to the lensed spectra to quantify the resulting variation in elemental mass-fraction estimates. 
We find that microlensing has a limited effect on SN Ia spectral and chromatic distortions but can introduce an additional uncertainty of $\sim$0.2 dex in elemental mass-fraction estimates. 
This microlensing uncertainty does not significantly affect the correlation between redshift and Ca abundance discussed in Section \ref{sec:results}. 

\begin{table*}[htb!]
    \centering
    \begin{tabular}{ccc}
        \hline
        $\kappa$ & $\gamma$ & $s$ \\
        \hline
        0.42     & 0.4      & 0   \\
        0.64     & 0.67     & 0   \\
        \hline
    \end{tabular}
    \caption{The list of microlensing magnification maps selected from the GERLUMPH dataset for this study. }
    \label{tab:lensparameter}
\end{table*}

\end{document}